\documentclass{article}
\font\sqi=cmssq8
\def\DR{\rm I\kern-1.45pt\rm R}
\def\DC{\kern2pt {\hbox{\sqi I}}\kern-4.2pt\rm C}
\newcommand{\beq}{\begin{equation}}
\newcommand{\eeq}{\end{equation}}
\newcommand{\bea}{\begin{eqnarray}}
\newcommand{\eea}{\end{eqnarray}}

\newcommand{\be}{\begin{equation}}\newcommand{\ee}{\end{equation}}

\newcommand{\p}[1]{(\ref{#1})}

\newcommand{\cD}{{\cal D}}
\newcommand{\cbD}{\overline{\cal D}}

\newcommand{\cZ}{{\cal Z}}
\newcommand{\cbZ}{\overline{\cal Z}}

\newcommand{\bQ}{{\overline Q}}

\begin{document}\begin{center}
{\bf \Large  ${\cal N}=8$ supersymmetric mechanics
on special K\"ahler manifolds}\\
\vspace{0.5 cm}
{\large Stefano Bellucci$^1$, Sergey Krivonos$^2$ and Armen Nersessian$^{3}$}\\

 \vspace{1cm}
$\;^1$ {\it INFN, Laboratori Nazionali di  Frascati, P.O. Box 13,
I-00044 Frascati, Italy}\\
$\;^2${\it Laboratory of Theoretical Physics, JINR, Dubna, 141980, Russia}\\
$\;^3$ {\it Yerevan State University, %A.Manoogian, 1, Yerevan,375025
and Yerevan Physics Institute, Yerevan, Armenia}\\
$\;\;${\it Artsakh State University, Stepanakert, Nagorny Karabakh,
Armenia}
\end{center}

\begin{abstract}
We propose  the  Hamiltonian model of ${\cal N}=8$ supersymmetric
mechanics on  $n-$dimensional special   K\"ahler manifolds (of
the rigid type).
\end{abstract}
\subsubsection*{Introduction}
One of the most studied theories since its introduction has been
the concept of supersymmetric mechanics \cite{witten}.
In spite of its simplicity this system
inherits the qualitative properties of  supersymmetric
field/string theories. Moreover, it is relevant to the specific
problems of condensed matter, quantum optics and mathematical
physics. The quantum-mechanical studies of this system were
focussed mainly on the simplest case of standard ${\cal N}=2$
supersymmetry\footnote{Hereafter  ${\cal N}$ denotes the number of
{\it real} supercharges. Often one considers supersymmetric
mechanics with $\cal N$ even, dividing the supercharges into
complex pairs $(Q,{\bar Q})$.} (see the review in \cite{sukh} and
refs therein). The systems with ${\cal N}=4$ supersymmetry also
received enough attention, but mostly in the classical context,
concentrated on  building  the appropriate Lagrangian  and
Hamiltonian  models of supersymmetric mechanics (see, e.g.,
\cite{leva,ikp,dpt,ikl1,korea,bn} and refs therein).

For many reasons, the most interesting case seems to be the ${\cal
N}=8$ supersymmetric mechanics. First of all, due to general
considerations developed in \cite{GR} this is the highest $\cal N$
case of {\it minimal} ${\cal N}$-extended supersymmetric mechanics
admitting realization on ${\cal N}$ bosons (physical and
auxiliary) and ${\cal N}$ fermions. Secondly, systems with eight
supercharges are the highest $\cal N$ ones among extended
supersymmetric systems which still possess a non-trivial geometry
in the bosonic sector \cite{WP}. When the number of supercharges
exceeds 8, the target spaces are restricted to be symmetric
spaces.
 Finally, ${\cal N}=8$ supersymmetric mechanics   should be related via
a proper dimensional reduction with four-dimensional  ${\cal N}=2$
supersymmetric field theories. So, one may hope that some
interesting properties of the latter
 will survive after reduction.
Nevertheless, the study of  ${\cal N}=8$ supersymmetric mechanics
remains in the initial stage, though there were a few interesting
investigations on this subject \cite{DE,gps,n8,abc}.
 Particulary,
 the ${\cal N}=8$ supersymmetric mechanics constructed by the
use of the vector supermultiplet reduced to one dimension at the
prepotential level has been considered in \cite{DE}. However, for the
reduced  ({\bf 5,8,3}) supermultiplet involving five physical
bosons, supersymmetric mechanics is quite different from
the original four-dimensional  ${\cal N}=2$ super Yang-Mills
theory. In \cite{abc}  the ${\cal N}=8$
supersymmetric mechanics  was considered
using the ({\bf 2,8,6}) supermultiplet
constructed by the dimensional reduction of four-dimensional
${\cal N}=2$ super Yang-Mills theory  at the level of the {\it field
strength}.

In this note we propose the Hamiltonian mechanics with ${\cal
N}=8$ supersymmetry algebra\footnote{We use the following
convention for the skew-symmetric tensor$\epsilon$ :
$\epsilon_{ij}\epsilon^{jk}=\delta_i^k,\;
\epsilon_{12}=\epsilon^{21}=1$ .} \beq
\{Q_{i\alpha},Q_{j\beta}\}=\{\bQ_{i\alpha},\bQ_{j\beta}\}= 0,\quad
\{Q_{i\alpha},\bQ_{j\beta}\}=
\epsilon_{\alpha\beta}\epsilon_{ij}{\cal H}_{SUSY},\qquad i,j\;
=\; 1,2;\;\alpha ,\beta =1, 2\;\; . \label{1}\eeq
 We require the configuration space of the underlying bosonic mechanics to be
a K\"ahler manifold $(M_0, g_{ab}dz^a d{\bar z}^b)$, as in the
${\cal N}=4$ mechanics considered in Ref. \cite{bn}. Our
model has the $(2n.4n)_{\DC}-$dimensional phase space, while in
its ${\cal N}=4$ counterpart the phase space is the
$(2n.2n)_{\DC}$-dimensional one. In the former system the
supercharges have  a cubic dependence
 on the fermionic variables, while in the latter one they have a linear dependence.
There is no doubt, that  our model is the Hamiltonian counterpart
of the supersymmetric mechanics, which could be constructed at the
Lagrangian level within the superfield approach by the use  of the
({\bf 2,8,6}) supermultiplet suggested in \cite{abc}.

 We find that ${\cal N}=8$ supersymmetry yields a strong restriction on the
geometry of configuration space: it has to be a {\sl special
K\"ahler manifold (of the rigid type)} (see, e.g. \cite{fre}). In
a distinguished coordinate frame its K\"ahler potential is given
by the expression
\begin{equation}
K(z,\bar z)= i\left( \frac{\partial f(z)}{\partial z^a}\bar z^a -
\frac{\partial \bar f(\bar z)}{\partial{\bar z}^a} z^a\right).
\label{2}
\end{equation}
During the last decade such a manifold was paid much attention to, due
to its relevance for the phenomenon of electric-magnetic
 duality  of  four-dimensional ${\cal N}=2$
  super Yang-Mills theories \cite{SW}.
The proposed  system inherits, besides the special  K\"ahler
geometry,  the duality symmetry of super Yang-Mills theory, which
makes the  constructed  ${\cal N}=8$ supersymmetric mechanics a
good ``probe'' to analyze some subtle properties of ${\cal N}=2,
d=4$ superYang-Mills theory. Finally, much like the
four-dimensional ${\cal N}=2$ super Yang-Mills case \cite{IZ} the
action of our supersymmetric mechanics can be modified by
introducing Fayet-Iliopoulos terms which give rise to potential
terms.

\subsubsection*{The model}
In order to construct ${\cal N}=8$ supersymmetric mechanics, let us
define the   $(2n.4n)_{\DC}$-dimensional symplectic structure
\begin{equation}
\Omega=d{\cal A}=d\pi_a\wedge dz^a+ d{\bar\pi}_{\bar a}\wedge d{\bar z}^{\bar a}
- R_{a\bar bc\bar d}\eta^c_{i\alpha}\bar\eta^{d|i\alpha}
dz^a\wedge d{\bar z}^{\bar b}+ g_{a\bar b}D\eta^a_{i\alpha} \wedge
D{\bar\eta}^{\bar b| i\alpha}\quad , \label{ss}\end{equation} where
\beq\label{a}
 {\cal A}=\pi_a dz^a+\bar\pi_{\bar a} d{\bar z}^{\bar a}
+\frac 12\eta^a_{i\alpha}g_{a\bar b}D\bar\eta^{\bar b\vert i\alpha}+ \frac
12\bar\eta^{\bar b}_{i\alpha}g_{a\bar b}D\eta^{a| i\alpha},\qquad
D\eta^a_{i\alpha} =d\eta^a_{i\alpha}+\Gamma^a_{bc}\eta^b_{i\alpha}
dz^c, \quad \eeq
 and
 $\Gamma^a_{bc}=g^{\bar d a}g_{b\bar d, c}$,
$ R_{a\bar bc\bar d}=-g_{e\bar b}(\Gamma^e_{ac})_{,\bar d}$ are,
respectively, the components of the connection and curvature of
the K\"ahler structure. The corresponding Poisson brackets are
given by the following non-zero relations (and their
complex-conjugates): \beq \{\pi_a, z^b\}=\delta^b_a,\quad
\{\pi_a,\eta^b_{i\alpha}\}=-\Gamma^b_{ac}\eta^c_{i\alpha},\quad
\{\pi_a,\bar\pi_{\bar b}\}= R_{a\bar bc\bar
d}\eta^c_{i\alpha}{\bar\eta}^{\bar d|i\alpha},
 \quad \{\eta^a_{i\alpha},
\bar\eta^{\bar b|j\beta}\}= g^{a\bar b}\delta_i^j\delta_\alpha^\beta.
\eeq
It is clear that the symplectic structure is {\sl covariant}
under following holomorphic transformations: \beq {\tilde z}^a=
{\tilde z}^a(z),\quad {\tilde\eta}^a_{i\alpha}=\frac{\partial
  {\tilde z}^a(z)}{\partial z^b}\eta^b_{i\alpha},
\quad{\tilde\pi}_a=\frac{\partial z^b}{\partial{\tilde z}^a}\pi_b
\;. \label{ht}\eeq Let us search for the supercharges among the
functions \beq Q_{i\alpha}=\pi_a\eta^a_{i\alpha}+\frac 13 {\bar
f}_{\bar a\bar b\bar c} {\bar T}^{\bar a\bar b\bar c}_{i\alpha},
\quad \bQ_{i\alpha}={\bar\pi}_a{\bar\eta}^{\bar a}_{i\alpha}+ \frac 13
f_{abc}T^{abc}_{i\alpha}.
%\Rightarrow\{Q^{\pm}_{i\alpha}, Q^{\pm}_{j\beta}\}=0,
\quad{\rm where}\quad T^{abc}_{i\alpha}\equiv
\eta^a_{i\beta}\eta^{b\vert j\beta}\eta^c_{j\alpha}\; .
\label{qqd}\eeq Calculating  mutual Poisson brackets of
$Q_{i\alpha},\bQ_{i\alpha}$
 one gets that they obey the
 ${\cal N}=8$ supersymmetry algebra (\ref{1}), if the following relations
 hold:
\beq \frac{\partial}{\partial{\bar z}^d}f_{abc}=0\; , \qquad
R_{a\bar b c\bar d}=- f_{ace}g^{e\bar e'}{\bar f}_{\bar e'\bar b\bar d}.
 \label{comp}\eeq
  The above equations  guarantee,
respectively, that  the first and the second
 equations in (\ref{1}) are fulfilled.
Then, we could immediately get the
  ${\cal N}=8$  supersymmetric  Hamiltonian
\beq
{\cal H}_{SUSY}=
%\frac 14\{Q^+_{i\alpha}, Q^{(-)i\alpha}\}=
\pi_a g^{a\bar b}\bar\pi_b+ \frac 13 f_{abc;d}\Lambda^{abc d} +\frac
13{\bar f}_{\bar a\bar b\bar c;\bar d}\bar\Lambda^{\bar a\bar b\bar c\bar d}
+ f_{abc}g^{c\bar c'}{\bar f}_{\bar c'
\bar d\bar e} \Lambda_0^{ab\bar d\bar e} \;, \eeq
where
\beq \Lambda^{abcd}\equiv -\frac 14
\eta^a_{i\alpha}\eta^{bi\beta}\eta^c_{k\beta}\eta^{dk\alpha},\quad
\Lambda_0^{ab\bar c\bar d}\equiv\frac 12
(\eta^{a\alpha}_i\eta^b_{j\alpha}\bar\eta^{\bar c\vert\beta i}
\bar\eta^{\bar dj}_\beta + \eta^{ai}_\alpha\eta^b_{i\beta}
\bar\eta^{\bar c\vert j\alpha} \bar\eta^{\bar d\beta}_{j}) \eeq
and the
covariant derivatives of the  third rank symmetric tensor  are defined as
$$f_{abc;d}=f_{abc,d}-
\Gamma^e_{da}f_{ebc}-\Gamma^e_{db}f_{aec}-\Gamma^e_{dc}f_{abe} \;.
$$
The equations (\ref{comp})  precisely  mean that the configuration space
$M_0$ is a {\sl special K\"ahler manifold  of the rigid type}\cite{fre}.
Taking into account the symmetry of $f_{abc}$ over spatial indices and
the
explicit expression of $R_{a\bar b c \bar d}$ via the metric $g_{a\bar  b}$,
we can  immediately  find the local solution for  the equations (\ref{comp})
\beq
f_{abc}=
\frac{\partial^3 f(z)}{\partial z^a\partial z^b\partial z^c},\qquad
g_{a\bar b}={\rm e}^{i\nu}\frac{\partial^2 f(z)}{\partial z^a\partial z^b}
+
{\rm e}^{-i\nu}
\frac{\partial^2 \bar f(\bar z)}{\partial \bar z^a\partial \bar z^b},\qquad
\nu=const\in \DR \;.
\label{gsk}
\eeq
Properly redefining the local function $f(z)$ and the odd coordinates $\eta^a_{i\alpha}$,
we shall get a supersymmetric mechanics  on the K\"ahler  space with metric   defined by the
 potential (\ref{2}).
For sure, this local solution is not covariant under an arbitrary holomorphic
transformation, and it assumes the choice of a distinguished
 coordinate frame.

The special K\"ahler manifolds of the rigid type
 became widely known   due to the so-called
``T-duality symmetry", which connects,  in the  of ${\cal N}=2,
 d=4$ super Yang-Mills theory,  the UV and IR limits of the theory \cite{SW}.
The ``T-duality symmetry''
 is expressed as follows:
\beq
\left( z^a,\; f(z)\right)\Rightarrow
\left( u_a=\frac{\partial f(z)}{\partial z^a},\; {\widetilde f}(u)\right),
\quad {\rm where}\quad
 \frac{\partial^2{\widetilde f}(u)}{\partial u_a\partial u_c}
\frac{\partial^2 f}{\partial z^c\partial z^b}=-\delta^a_b\; .
\label{duality}\eeq
By the use of (\ref{ht}), we can extend the duality transformation
(\ref{duality})  to that  of the
whole phase superspace $
(\pi_a,\,  z^a ,\, \eta^a_{i\alpha})\to (p^a, u_a, \psi_{a|i\alpha})$
\beq
 {u}_a=\partial_a f(z),\quad
p^a\frac{\partial^2 f}{\partial z^a\partial z^b}=-\pi_b
 ,\quad \psi^{i\alpha}_a=
\frac{\partial^2 f}{\partial z^a\partial z^b}\eta^{b|i\alpha}.
\eeq
Now, taking into account the expression of the symplectic structure in terms of the
presymplectic one-form (\ref{a}), we can easily perform the Legendre
transformation of the Hamiltonian to the (second-order) Lagrangian
\bea
&&{\cal L}= {\cal A}(d/dt)-{\cal H}_{SUSY}\vert_{\pi_a=g_{a\bar b}
{\dot{\bar z}}^b}
= \nonumber\\
&&=g_{a\bar b}{\dot z}^a{\dot {\bar z}}^{\bar b}
+\frac 12\eta^a_{i\alpha}g_{a\bar b}\frac{D\bar\eta^{\bar b|i\alpha}}{dt}+
\frac 12 \bar\eta^{\bar b}_{i\alpha}g_{a\bar  b}\frac{D\eta^{a|i\alpha}}{dt}-
\frac 13
f_{abc;d}\Lambda^{abcd}
-\frac 13{\bar f}_{\bar a\bar b\bar c;\bar d}\bar\Lambda^{\bar a\bar b\bar c\bar d} -
f_{abc}g^{c\bar c'}{\bar f}_{\bar c' \bar d\bar e}\Lambda_0^{ab\bar d\bar e}
.
\label{lag}\eea
Here we denoted ${d/dt}={\dot z}^a\partial/\partial z^a+
\dot\eta^a_{i\alpha}\partial/\partial\eta^a_{i\alpha} +c.c.\;$.
Clearly, the  Lagrangian (\ref{lag})
is covariant under holomorphic transformations (\ref{ht}),
 and duality transformations as well.
As we noticed in Introduction, the above
 Lagrangian admits the superfield description
based on the ({\bf 2, 8, 6}) ${\cal N}=8, d=1$ supermultiplet
constructed in \cite{abc}. The basic object is the ${\cal N}=8$ complex
superfield $\cZ, \cbZ$ subject to the constraints
\bea
&& \cD^{ia}\cZ=0 ,\quad \cbD^{ia}\cbZ=0, \label{con1} \\
&& \cD^{i(a}\cD^{jb)}\cbZ+\cbD^{i(a}\cbD^{j b)}\cZ=0,\;\quad
\cD^{(i a}\cD^{j)b}\cbZ-\cbD^{(i a}\cbD^{j)b}\cZ=0,  \label{con2}
\eea
where the spinor derivatives obey the following relations:
\be\label{4bder}
\left\{ \cD^{ia},\cD^{jb}\right\}=\left\{ \cbD^{ia},\cbD^{jb}\right\}=0,\quad
\left\{ \cD^{ia},\cbD^{jb}\right\}=2i\epsilon^{ij}\epsilon^{ab}\partial_t\,.
\ee
The equations \p{con1},(\ref{con2}) represent the direct reduction of the
constraints describing the ${\cal N}=2,d=4$ vector multiplet.
The most general ${\cal N}=8$ superfield action mimics the
action for the ${\cal N}=2,d=4$ case
\be
{ S}= {\rm Im}\;\int dt d^4\theta f(\cZ) \;.
\label{saction1}
\ee
Being integrated over $\theta,\bar\theta$, exploiting the constraints
\p{con1},\p{con2} and eliminating the auxiliary fields
by their equations of motion, the action \p{saction1} coincides with
the one defined by the Lagrangian \p{lag}.
Thus, we see that our system
indeed is closely related
with the action for the ${\cal N}=2, d=4$ vector supermultiplet.
Clearly enough, the special
K\"ahler geometry arises naturally from the action as in the $d=4$ case.
The analogy with
$d=4$ goes even further, since  the proposed   system
exhibits the Seiberg-Witten duality. Moreover, one can  modify the
 actions by adding
two possible Fayet-Iliopoulos terms with some arbitrary constant vectors
 as it has been done in $d=4$ \cite{IZ}.
For a special choice of the superpotential, we expect that the system  will have
also conformal symmetry.
The detailed discussion
of these cases
goes beyond the scope of the present paper and will be carried
 out elsewhere.
\\

{\bf Acknowledgments.}
 The work was supported in part by
the European Community's Human Potential Programme under contract
HPRN-CT-2000-00131 Quantum Spacetime and
the NATO Collaborative Linkage Grant PST.CLG.979389 (S.B.),
RFBR-DFG grant No 02-02-04002, grant DFG No 436 RUS 113/669, and RFBR grant
No 03-02-17440 (S.K.),
the INTAS-00-00254 (S.B. and S.K.)  and  INTAS 00-00262 (A.N.) grants.
S.K. thanks INFN --- Laboratori Nazionali di Frascati  for the warm
hospitality extended to him during the course of this work.

\end{document}